**Author's Post-Print (final draft post-refereeing)**

# Boundary node detection and unfolding of complex non-convex ad hoc networks

SE-HANG CHEONG, University of Macau
YAIN-WHAR SI, University of Macau

Complex non-convex ad hoc networks (CNCAH) contain intersecting polygons and edges. In many instances, the layouts of these networks are also not entirely convex in shape. In this paper, we propose a Kamada-Kawai based algorithm called W-KK-MS for boundary node detection problem, which is capable of aligning node positions while achieving high sensitivity, specificity and accuracy in producing a visual drawing from the input network topology. The algorithm put forward in this paper selects and assigns weights to top-$k$ nodes in each iteration in order to speed up the updating process of nodes. We also propose a novel approach to detect and unfold stacked regions in CNCAH networks. Experimental results show that the proposed algorithms can achieve fast convergence on boundary node detection in CNCAH networks and are able to successfully unfold stacked regions. The design and implementation of a prototype system called ELnet for analyzing CNCAH networks is also described in this paper. The ELnet system is capable of generating synthetic networks for testing, integrating with force-directed algorithms, and visualizing and analyzing of algorithms' outcomes.

## 1. INTRODUCTION

Advances in low-power and miniaturization design have had tremendous impact on the development of useful equipment and services. These new services and devices can be used for high-end applications as well as for basic consumer oriented products. Most of these devices have built-in wireless modules, even accelerometers and GPS (Global Positioning System) features. Moreover, development of advanced protocols in wireless communications and rapid advancement in electronic technologies enable the large scale deployment of wireless devices in ad hoc networks. To this end, localization problems in wireless and ad hoc networks formed by these devises become an important research topic.

Force-directed algorithms are frequently used in network visualization. Force-directed algorithms rely on spring forces. Forces between the nodes can be computed based on their graph theoretic distances, determined by the lengths of shortest paths between them. There are repulsive forces between all nodes, but also attractive forces between nodes that are adjacent. Force-directed algorithms often define an objective function. A layout for a graph is then calculated by finding a minimum of this objective function in which adjacent nodes are near from each other, and non-adjacent nodes are well spaced. Force-directed algorithms calculate the layout of a graph using only information contained within the structure of the graph itself. Graphs drawn with these algorithms tend to exhibit symmetries, and produce crossing-free layouts for planar graphs [Tamassia 2007].

They can be used to produce a visual drawing that is proportional to the given network topology by using the spring force exerted on nodes and edges. These algorithms include Kamada-Kawai [4], Fruchterman Reingold [5], and Davidson Harel [6]. The objective of the Kamada-Kawai algorithm is to draw a graph that is as planar as possible. The Kamada-Kawai algorithm uses only the information of nodes and edges when it produces visual drawings from the network topologies. The Kamada-Kawai algorithm adjusts the positions of nodes iteratively in order to achieve a state of equilibrium. It uses an energy function to represent the state of equilibrium.

In this paper, we propose an algorithm called W-KK-MS for boundary node detection which uses a batch weight updating algorithm and signal strength for node adjustment. We target the localization problems associated with a particular type of network called Complex Non-Convex Ad Hoc (CNCAH) Networks. A CNCAH network comprises of intersecting edges and complex polygons (i.e. convex and non-convex polygons). The layouts of CNCAH networks are usually not entirely convex in shape [Bresenham et al. 1991].

Unfolding and boundary node detection are two major problem areas of CNCAH. One particular example involving CNCAH networks is an ad hoc emergency network which is designed for use in situations such as earthquakes or tsunami [Cheong et al. 2011]. Most of the existing telecommunication networks are not designed to be fault-tolerant and back-up systems are often not available during emergencies. For instance, these telecommunication networks could be partially destroyed or interrupted during natural disasters. This is often due to the damage caused to the stations or the fact that networks become overwhelmed by sudden transmission spikes in an effected area. During or after disasters, victims who become trapped in the disaster areas or under debris could establish an ad hoc mobile network by using their hand-held devices and might attempt to communicate with the rescuers. Therefore, capabilities for determining node localization information and estimating the network layout can be extremely helpful for the rescue teams. In this paper, we propose an algorithm called W-KK-MS for boundary node detection which uses a batch weight updating algorithm and signal strength for node adjustment. The W-KK-MS is an extended version of Kamada-Kawai (KK) algorithm and it can be applied to CNCAH network topologies. Moreover, the proposed algorithm is also designed to unfold stacked regions in visual drawings. Localization is a process of estimating the position of a node in a network according to some spatial coordinate system [Perkins et al. 2005]. Locations of individual nodes in practical applications such as highway, battlefield, and logistics are used in routing protocols and network coverage analysis [Virrankoski 2003].

Boundary node detection is important in wireless sensor networks which comprise of a set of small devices (nodes with tiny sensors) for applications such as sensing, monitoring, and data collection. These nodes often form an ad hoc network to collect data and report to a designated station. Coverage and connectivity are two important aspects of a wireless sensor network. Any failure in the network caused by the deployment, disconnection, or sensor failure can be analyzed by detecting the boundary nodes. Such detection can reveal the status of the coverage and connectivity of a given network [Zhang et al. 2006].

Unfolding is a process to open up twisted and stacked regions of the visual drawing. Figure 1 illustrates the folded and unfolded regions of a visual drawing. Folded and unfolded regions in the visual drawing are highlighted with light blue color. Unfolding is important in network localization because it provides a valuable insight into the structure and layout of networks. Unfolding is useful for large complex networks that are groups of nodes, densely connected within and only loosely connected with the rest of the network [Blondel et al. 2008; Subelj and Bajec 2011; Wickramaarachchi et al. 2014].

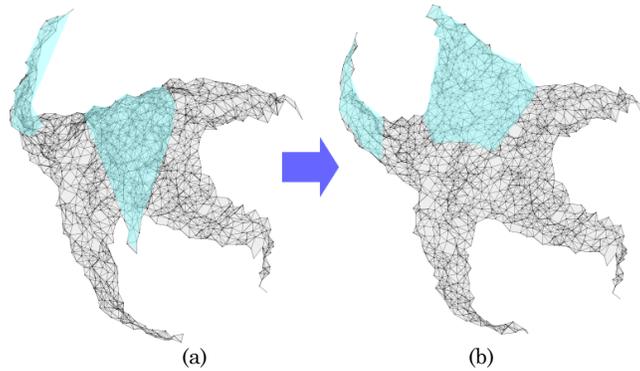

(a) (b)

Figure 1 (a) Folded regions of a visual drawing, (b) Unfolded regions of a visual drawing.

There are a number of systems and software libraries such as NetworkX [NetworkX developer team 2014], Open Graph Drawing Framework [Gutwenger et al. 2013], and JGraphT [Naveh 2013] available for generating simple networks and analyzing relevant properties. However, these systems cannot be easily tailored for analyzing CNCAH networks. For instance, these applications cannot be easily extended for generating CNCAH networks. Besides, the functions for integrating with third party force directed algorithms, and visualization and analysis of these algorithms' outcomes are not supported. Against this background, we developed a prototype system called ELnet which is capable of generating CNCAH network topologies for experiments, simulating various force-directed algorithms for evaluating their performance, and visualizing the algorithms' performance.

## 2. CONTRIBUTION

The main objective of the paper is to find boundary nodes and unfold twisted and folded regions in CNCAH networks. We also developed a prototype system called ELnet for generating CNCAH networks and evaluating algorithms in this paper. For example, ELnet generates CNCAH networks with combinations of arbitrary geometric shapes for our experiments. The scientific and technical contributions of this work are twofold.

1. First, we extend the Kamada-Kawai (KK) algorithm for boundary node detection and unfolding of CNCAH networks. Generally, the KK algorithm is slow for large and complex networks [Nooy et al. 2005]. In this paper, we propose an algorithm called W-KK-MS, which is capable of adjusting the node positions while maintaining an acceptable level of sensitivity, specificity and accuracy in not only efficient in producing a visual drawing from the input network topology. W-KK-MS algorithm is but also capable of discovering folded and twisted regions and minimize edge crossings in CNCAH networks. The proposed W-KK-MS algorithm considers signal strength in producing visual drawings. W-KK-MS not only minimizes the edge crossings, but also significantly improves the performance for locating boundary nodes. Therefore, the proposed algorithm is more appropriate for wireless sensor networks than KK. The visual drawing is then used for detecting boundary nodes from the input network topology. Moreover, for complex network topologies without anchor information, the quality of visual drawings produced by force directed algorithms such as KK can be poor for CNCAH networks [Efrat et al. 2010]. For instance, some parts of the network may be folded or twisted. For this reason, the proposed W-KK-MS algorithm is also designed to discover and unfold twisted regions. Experimental results show that the proposed algorithm can achieve fast convergence on boundary detection in CNACH networks and area able to successfully unfold stacked regions.

2. Next, the design and implementation of a prototype system called ELnet for analyzing CNCAH networks is presented. To the best of authors' knowledge, existing tools are incapable of generating CNCAH networks for testing. In this paper, we describe a prototype system which is capable of generating synthetic networks, integrating with third party algorithms, visualizing experiment outcomes and performance. As part of the system, we also develop a boundary node identification algorithm which can be used to discover nodes on outward boundary of a 2-dimensional plane. That is, after the visual drawing for a network topology is projected onto a geometric plane using force directed algorithms, we can use the boundary node identification algorithm to identify the true boundary nodes. Once the true boundary nodes are identified, we can compare them with the results obtained from the force directed algorithms. The graphical interface of ELnet also provides features for interactive visualization of boundary nodes from a visual drawing and functions to plot the results of the analysis.

In Section 3, we summarise existing well-known force-directed methods and the recent studies on boundary node detection problem. In Section 4, we propose an algorithm called W-KK-MS for detecting boundary nodes and unfolding stacked regions in CNCAH networks. In Section 5, we briefly introduce the design of a prototype system called ELnet for analyzing CNCAH networks. In Section 6, we evaluate the performance of the proposed algorithm. In Section 7, we conclude the paper and discuss the future work.

**3. RELATED WORK**

Force-directed algorithms are used in many application areas. They are especially useful for visualizing graphs from social networks [Quinn and Breuer 1979]. Force-directed algorithms such as Kamada-Kawai (KK) [Quinn and Breuer 1979] and Fruchterman Reingold (FR) [Fruchterman and Reingold 1991] are based on Eades's spring-embedder model. These two approaches attempt to minimize the edge crossing and distribute nodes and edges uniformly. A similar algorithm was also proposed by Davidson Harel (DH) [Davidson and Harel 1996], which is based on a simulated annealing process for drawing of graphs.

Among these force-directed algorithms, KK focuses on the relations among the nodes from the entire network topology. It calculates the value of the energy remaining on every node, and then iteratively adjusts the nodes with highest energy until all nodes reach their minimum energy. An equilibrium state is obtained when the energy on the nodes is close or equal to 0. However, nodes in the visual drawing can be stacked together when force-directed algorithms such as KK is used for display. These algorithms generally cannot guarantee a planar result especially for large and complex networks. Furthermore, some parts of the network could be twisted in the final outcome. For example, a u-shaped network topology could be twisted into a w-shape or a z-shape [Kevin et al. 2010]. There are exiting studies using modularity to detect stacked nodes in the network topologies. Modularity is an attribute which specifies the distribution of nodes in a specified region. That is, if the modularity of a region is high, nodes in this region may be stacked [Alex et al. 2007; Blondel et al. 2008]. However, this assumption may not be always true for complex network topologies or network topologies with clusters. Since the distribution of nodes is non-uniform in complex non-convex network topologies. Nodes with high modularity may occur on the gap rather than on the boundary of clusters. In contrast to their approaches, our approach keeps track the changes of a visual drawing iteratively. Specifically, our algorithm locates the regions which are significantly different from the initial network topology during the execution. When these regions are identified, the algorithm then attempts to repair these regions.

In previous studies about the localization of ad hoc networks, force-directed algorithms are used to perform position tracking to improve data flow in the network. Efrat et al. [Efrat et al. 2010] applied a multi-scale dead-reckoning algorithm for sensor localization in force-directed algorithms by using the length of edges and angular information. The performance of their approach was evaluated on non-convex network topologies. A multi-scale algorithm usually uses multiple models at different scale to resolve a large scale of problem. After dividing a large network into smaller sub-networks, a multi-scale algorithm uses several level of stages to process these sub-networks. Specifically, sub-networks are processed in stages with local organization schema and each stage computes one or more sub-networks. Multi-scale algorithms can also compute some of the stages simultaneously and the results of stages are incrementally combined until all sub-networks are computed [Barth et al. 2012; Weinan 2011]. Efrat et al. also use the dead reckoning for estimating the distance of nodes. Dead reckoning is an updating process of the current positions of nodes by using previously calculated positions or estimated speed over elapsed time and course. The problem of detecting boundary nodes and holes in hoc networks has been widely reported in literature. Dong et al. [Dong et al. 2009] proposed an algorithm for detecting holes in the network by using fine-grained boundary recognition. This algorithm can be used to detect inner and outer boundary cycles using the information of nodes and edges of the network. The algorithm relies on global connectivity information of network in which shortest path trees of primary boundary circle are used on the boundary refinements [Liu et al. 2012]. Wang et al. [Wang et al. 2006] proposed an algorithm to detect the boundaries of holes using the shortest path tree. In the proposed algorithm, distinct portions of similar paths that span the network are selected for detection [Li and Liu 2010]. The assumption of the proposed algorithm is that if there is not a hole between the nodes within the shortest path tree, the parts are more similar to straight lines. The proposed algorithm builds the shortest path tree by flooding the network from an arbitrary root node upon initialisation. Blondel et al. [Blondel et al. 2008] proposed an algorithm to unfold communities in large scale of social networks. This algorithm uses heuristic method that unfold the community structures of social networks based on the modularity optimization. That is, if the modularity of a region is high, the region may be folded. Wickramaarachchi et al. [Wickramaarachchi et al. 2014] proposed an algorithm for unfolding communities in large graphs by using a greedy modularity maximization approach. The algorithm is designed for parallel computing, unfolding communities, and minimizing the cross edges between folded communities in large graphs.

Volker et al. [Volker et al. 2012] also proposed an approach for tracking the position of nodes using a force-directed algorithm based on signal strength and step recognition. Step recognition is a process to collect the movement of sensors. Their approach uses movement information to adjust the assignment of force in the force-directed algorithms. That is, they use a built-in accelerometer which is installed on sensors to identify the step status of sensors so that they can collect the information of movement from sensors. Their approach was evaluated by using experimental data obtained from 60-device wireless sensor network deployed in two buildings. They evaluated the influence of position and errors of estimation with anchor points (the position of 60 devices is known) verses different kinds of distance estimation methods. In contrast to their approaches, our approach extends Kamada-Kawai by using a batch weight updating algorithm to guide the movement of nodes and edges during the execution. One of the key contributions of our approach is the ability to process CNCAH networks without any location information except the topology.

Graph generators such as NetworkX [NetworkX developer team 2014], Open Graph Drawing Framewor [Gutwenger et al. 2013], and JGraphT [Naveh 2013] are commonly used for generating network topologies. There are also tools and simulators available for visualizing wireless sensor networks [Buschmann et al. 2005; Österlind et al. 2010; Shu et al. 2008]. However, these tools are not specially designed for the evaluation of force-directed algorithms for boundary node detection in ad hoc networks.

**4. WEIGHTED KAMADA-KAWAI WITH MULTI-NODE SELECTION**

Kamada-Kawai (KK) [Kamada and Kawai 1989] is a visualization algorithm that is based on Eades's spring-embedder model [Eades 1984]. The objective of the algorithm is to distribute nodes and edges uniformly and minimize edge crossing [Chen 2006]. The key idea behind this algorithm is to use an energy function to model the spring on network topologies. The energy function $E$ used in KK is described in equation 1.

$$E = \sum_{i=1}^{n-1} \sum_{j=i+1}^{n} \frac{1}{2} k_{i,j} (|p_i - p_j| - l_{i,j})^2 \tag{1}$$

The above energy function is used to calculate a visual position for nodes in the network topology so that their visual distance is proportional to their theoretical graphed distance. In the energy function, $k_{i,j}$ is the stiffness of the spring of node $i$ and $j$, $p_i$ and $p_j$ are the visual positions of node $i$ and $j$, and $l_{i,j}$ is the theoretical graphed distance of node $i$ and $j$. The theoretical graphed distance of the spring ($l_{i,j}$) between node $i$ and $j$ can be defined as follows:

$$l_{i,j} = \frac{L_0}{\max_{i<j} d_{i,j}} \times d_{i,j} \tag{2}$$

where $d_{i,j}$ represents the hop count between node $i$ and $j$. $d_{i,j}$ is the shortest hop count of the path among all possible paths between node $i$ and $j$. $L_0$ is the side length of the drawing frame, if the drawing frame is a rectangle, the longest side of the rectangle is chosen as $L_0$. In addition, the stiffness of the spring of node $i$ and $j$ can be calculated by:

$$k_{i,j} = \frac{K}{d_{i,j}^2} \tag{3}$$

where $K$ is a constant for scaling.

KK iteratively updates the visual position of nodes by using the Newton-Raphson method. For every iteration, the algorithm selects a node that has the highest value of energy remaining and updates its visual position in order to minimize the energy function $E$. From our preliminary testing, we find that KK is able to produce visual drawings of simple graphs with relatively fast convergence rate. This result is expected since the edge lengths in simple networks are not usually restricted by constraints. Hence, the length of the edges can vary significantly in a simple graph. However, for ad hoc networks in general and CNCAH networks in particular, the lengths of the edges are often constrained by the hardware limitation and signal strength. Therefore, in this article, we propose a number of extensions to the KK algorithm for unfolding and boundary node detection in CNCAH networks. The proposed algorithm called W-KK-MS is described in Algorithm 1 and an illustration of unfolding is depicted in Figure 2.

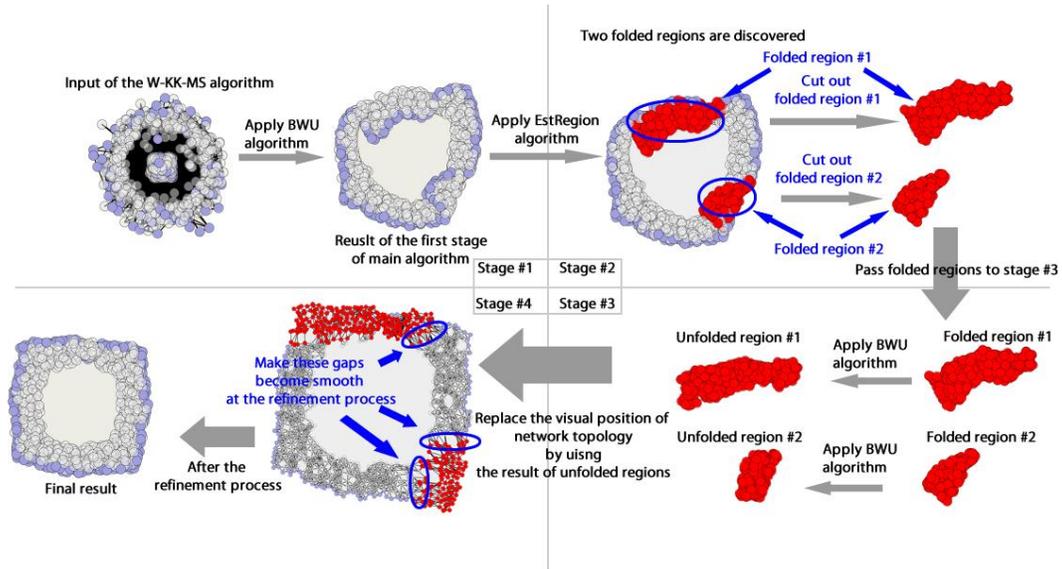

Figure 2 Illustration of unfolding by the W-KK-MS algorithm.

The W-KK-MS algorithm has four stages. The first stage is a batch weight updating approach for node movement which is designed to achieve fast convergence rate on boundary node detection. An array $w$ stores the weights of the nodes. The size of the array equals to the number of nodes $n$. For example, if the networks has 1000 nodes, then the size of the array $w$ is 1000. The initial values of the weights are set to 1. The second stage uses EstRegion algorithm to discover possible twisted and folded regions. The third stage is to unfold twisted and folded regions found in the second stage. The final stage is the substitution of visual position of nodes and refinement of the network.

In the first stage, the W-KK-MS algorithm uses Batch Weight Updating (BWU) algorithm to iteratively update the visual position of nodes. The iterative updating is controlled by an input parameter $P$ which is the percentage of the nodes designated for selection and node count $n$ of the network. That is, the first stage of main algorithm will terminate when $2 \times P \times n$ nodes in the network have been updated. For example, if a network has 1000 nodes and $P$ is 5%, the first stage of main algorithm will terminate when $2 \times P \times n$ (i.e. $2 \times 0.05 \times 1000 = 100$) nodes have been updated. For each iteration, the W-KK-MS algorithm first executes CalWeight algorithm to determine the weights of the nodes. Weights are assigned to every node in the network topology to influence the node selection process of the algorithm. The smaller the weight, the smaller the chance the node will be selected for visual position updating. On the contrary, the higher the weight, the higher the chance the node will be selected. Next, the BWU algorithm is executed to update the visual position of the nodes by using the weights determined by CalWeight algorithm. CalWeight and BWU are executed repeatedly until $2 \times P \times n$ nodes in the network have been updated. The details of CalWeight and BWU algorithms are explained in section 4.1 and 4.2.

In the second stage, EstRegion algorithm is first executed to find folded regions in the network. A number of strategies are also used in our implementation to repair folded regions from the network. The inner working of EstRegion is explained in section 4.3.

At the third stage, the unfolding process is performed. Firstly, the W-KK-MS algorithm selects one of the folded regions found in the second stage to initiate the unfolding process. Secondly, the W-KK-MS algorithm adds additional edges to some of the nodes in the folded region where the density of edges is low. Unfolding process may fail if there are insufficient edge connections among the folded regions. This concept is analogous to the case of two large clusters with few edge connections in between. In these situations, force-directed algorithms are not able to unfold these regions effectively because their edge connections are weak. Therefore, additional edges are added to create stronger edge connections among folded regions. Edges will be added to the gaps of the folded regions if the average degree of folded regions (illustrated in Figure 3) is less than a threshold $\varepsilon$.

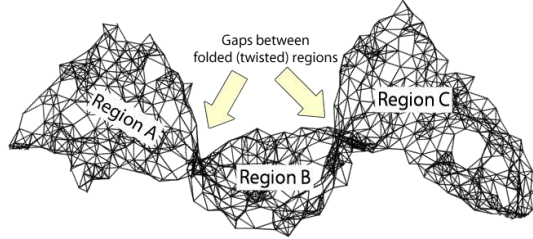

Figure 3 Gaps between folded regions.

Thirdly, the W-KK-MS algorithm assigns weights to the nodes from the folded region by using CalWeight algorithm. Next, the W-KK-MS algorithm uses BWU algorithm to update the visual position of nodes in the selected region with new assignment of weights so that distorted regions in the visual drawing can be unfolded. Finally, if more than one folded regions is found by the EstRegion algorithm in second stage of the W-KK-MS algorithm, the above steps are repeated for each folded region until all folded regions have been unfolded.

From the result of the second stage of W-KK-MS algorithm, folded regions can be discovered from the visual drawing of network topology. In the W-KK-MS algorithm, folded regions formed by a selection of nodes will be stored in an array called $R_{FOLDED}$ and the visual position of these nodes are all cloned from $G$. For example, a folded region is discovered by EstRegion algorithm at the second stage of W-KK-MS algorithm and the region is formed by three nodes $a(x_0, y_0)$, $b(x_1, y_1)$ and $c(x_2, y_2)$. The process of cloning is to copy the visual positions $(x, y)$ of the nodes from $G$ and insert them into $R_{FOLDED}$. That means, there are two identical copies of node $a$, $b$ and $c$ after the process of cloning. One copy of nodes is in $G$ and another copy is in $R_{FOLDED}$.

Next, the unfolding process is applied to $R_{FOLDED}$ at the third stage of W-KK-MS algorithm. The resulting unfolded regions (nodes with new visual positions assigned) are stored in another array called $R_{UNFOLDED}$ by using the cloning process similar to previous one. After that, the array $R_{UNFOLDED}$ and nodes of $G$ will be passed to the next stage. $R_{FOLDED}$ is discarded after the unfolding process. The last stage of W-KK-MS algorithm is to update (substitute) the visual position of the nodes in network topology $G$ by using the result of unfolding process ($R_{UNFOLDED}$) except for unfolded regions which still have high approximation error $r$ after the unfolding process. Because $R_{UNFOLDED}$ is the output of unfolding process and $R_{FOLDED}$ is the input of unfolding process, both arrays contain the same nodes with different visual positions. That is the reason why the W-KK-MS algorithm needs to replace the visual position of nodes in $G$ by using the visual position of nodes in $R_{UNFOLDED}$. Finally, the main algorithm will call BWU algorithm to refine the network topology $G$.

**ALGORITHM 1.** W-KK-MS algorithm

**Input**: a visual drawing of netwotk topology $\mathbf{G}(V, E)$

        percentage of node for selection **P**
**Output**: a visual drawing of network topology **G**

**Initialize** the iteration of algorithm **counter** ← 1;
**Initialize** the array of weight **w** ← ∅;
**Initialize** the node count of G **n** ← |V|;
**Initialize** the array containing the length of edges (before first stage) of G as $e_{INIT}$;

/∗ First stage.∗/
/∗ Heuristics based approach on node movements.∗/
**Repeat** until counter ≥ P × 2 × n
        w ← CalWeight(G, P, n, counter);
        G ← BWU(G, w, P);
        counter ← counter + 1;
**End Repeat**

/∗ Second stage.∗/
/∗ Find twisted and folded regions in G.∗/
**Initialize** the array containing the length of edges (after first stage) of G as $e_{UPDATED}$;
**Initialize** the threshold of the estimation of folded regions as **θ**;
**Initialize** the array of folded regions $R_{FOLDED}$ ← EstRegion( $e_{INIT}$, $e_{UPDATED}$, $\theta$ );

/∗ Third stage.∗/
/∗ Unfold twisted and folded regions.∗/
**Initialize** the array of unfolded regions $R_{UNFOLDED}$ ← ∅;
**Initialize** the threshold of adding new edges as **ε**;
counter ← 1;
**Foreach** R ∈ $R_{FOLDED}$ **Do**
        **Initialize** the node count of R as **n´**;
        **Initialize** the array containing edges of R as $R_E$;
        **Repeat** until counter ≥ P × 2 × n′ {
                w ← CalWeight(R, P, n′, counter);
                $E_1$ ← the value of energy of R; /∗ Before the position update.∗/
                R ← BWU(R, w, P);
                $E_2$ ← the value of energy of R; /∗ After the position update.∗/

                /∗ Add additional edges to nodes of folded region
                which have low average degree.∗/
                **If** $E_2 - E_1 < \varepsilon$ **Then**
                      /∗ v1 and v2 are nodes of R.∗/
                      **Foreach** v1 ∈ nodes of R **Do**
                          **Foreach** v2 ∈ nodes of R **Do**
                              /∗ Add an edge between v1 and v2.∗/
                              **If** hopcount(v1, v2) > avgdeg(v1, v2) **Then**
                                  $R_E$ ←< v1, v2 >;
                              **End if**
                          **End Foreach**
                      **End Foreach**
                **End If**

                /∗ Increase the counter.∗/
                counter ← counter + 1;
        **End Repeat**

```
        /∗ Append unfolded region to R_UNFOLDED.∗/
        R_UNFOLDED ← R;
End Foreach

/∗ Forth stage.∗/
/∗ Substitute visual position from R_UNFOLDED to G.∗/
Foreach R′ ∈ R_UNFOLDED Do
        /∗ Calculate the approximation error of edges of the unfolded region.∗/
        initialize the array of approximation error ERR ← ∅;
        Foreach e ∈ R_UNFOLDED Do
                ERR ← r(e); /∗ See section 4.3 for detailed description.∗/
        End Foreach
        Sort ERR in descending order;

        /∗ Ignore the unfolded region which still has high approximation error.∗/
        Foreach err ∈ ERR Do
                If  err > θ  Then
                        Delete R′;
                End If
        End Foreach
        If R′ is deleted Then
                break;
        End If

        /∗ Update the visual position of nodes in G
        by using the positoins from unfoled region R′.∗/
        Foreach v ∈ R′  Do
                G_v. x ← v. x;  /∗ Substitute the x position from R′ to G.∗/
                G_v. y ← v. y;  /∗ Substitute the y position from R′ to G.∗/
        End Foreach
End Foreach

/∗ Refine G by using BWU algorithm after visual position substitution.∗/
w ← CalWeight(G, P, n, 1);
BWU(G, w, P);
```

Algorithm 1 Pseudo code of main algorithm

**4.1 Batch Weight Updating (BWU) algorithm**

From our extensive testing, we find that KK updates the visual position of a node at each iteration and the updating process can be extremely slow. This is especially noticeable for large network topologies. Moreover, we also observe that nodes that are too close to each other within a particular region could have similar properties. They are more likely to have common neighbors and share similar signal strength. To make use of this property and to speed up the updating process, we extend the KK algorithm by selecting multiple nodes instead of a single node in each iteration. Specifically, the algorithm pushes the top-$k$ nodes that have a larger *change* ($\Delta$) onto an ordered stack. Next, the top-$k$ nodes from the ordered stack are popped out from the stack and their visual positions are updated. The value of *change* ($\Delta$) for these nodes is then recalculated again at the next iteration.

In addition, weights are also assigned to the selected nodes in each iteration. The values of weights assigned to these nodes not only reflect the magnitudes of updating visual position but also determines which nodes should be updated. For example, nodes within a closed region and have similar neighborhood are likely to be adjusted. Because these nodes have roughly the same value of average degree and therefore their positions should be updated in batches. Specifically, we can use weights to control the visual position update of nodes in some selected regions of the visual drawing as well as to keep other regions unchanged. The calculation of *change* ($\Delta$) for a node $i$ is given in equation (4). According to equation (4), weights are related to *change* ($\Delta$). The larger the weight, the higher the value of change. The higher the change a node has, the higher the chance the node will be selected for visual position update.

$$\Delta_i = \sqrt{\left(\sum_{j=1}^{n} w_i \times k_{i,j}\left(1 - \frac{l_{i,j}}{p_{i,j}}\right) \times x_i\right)^2 + \left(\sum_{j=1}^{n} w_i \times k_{i,j}\left(1 - \frac{l_{i,j}}{p_{i,j}}\right) \times y_i\right)^2} \quad (4)$$

where $w_i$ is the weight for node $i$, $p_{i,j}$ is the distance of node $i$ and $j$ visual positions. $l_{i,j}$ is the theoretical graphed distance of node $i$ and $j$. $x_i$ and $y_i$ are x and y coordinates of the visual position of node $i$. The estimated distance used in this paper is based on the assumption of free-space path loss (FSPL) [Goldsmith 2005]. According to this assumption, the estimated distance is calculated based on the signal strength of the nodes. FSPL is a term used in telecommunication to denote the loss in signal strength of an electromagnetic wave as a result of a line-of-sight path. The estimated distance (in meters) of FSPL can be calculated as follows:

$$d = 10^{\frac{27.55 - (20 \times \log_{10}(f)^{-s})}{20}} \quad (5)$$

where $f$ is the signal strength and $s$ is the signal frequency in Mhz.

The pseudo code of the Batch Weight Updating (BWU) algorithm is described in Algorithm 2. On initializaiton, the algorithm calcualtes the maximum change of every node by using equation (4) and stores them into an array $DM$. The algorithm then sorts the array $DM$ in descending order. Once the array $DM$ has been sorted, the algorithm selects top K nodes stored in $DM$ having high value of energy remaining for the next step.

The algorithm uses a nested loop which includes an outer and an inner loop to update the visual positions of the selected nodes. The visual position of selected nodes are updated by the inner loop which is also responsible for minimizing the energy function. The outer loop controls the iteration of inner loop. BWU will terminate when the count of outer loop is greater than the ratio of $V$ (node count) and $K$ (percentage of nodes for selection), i.e $\frac{|V|}{K}$.

---

**ALGORITHM 2.**   BWU algorithm

**Input**:  a visual drawing of netwotk topology **G**(V, E)
          a weight matrix **w** of nodes of network topology G
          percentage of node for selection **K**
**Output**: a visual drawing of network topology **G**

/∗ Calculate the value of energy remaining ofnodes in G.∗/
**Initialize** an array containing the value of energy remaining ofnodes as **DM**;
**Foreach** i ∈ V **Do**

$$DM[i] \leftarrow \sqrt{\left(\sum_{j=1}^{n} w_i \times k_{i,j} \left(1 - \frac{l_{i,j}}{p_{i,j}}\right) \times x_i\right)^2 + \left(\sum_{j=1}^{n} w_i \times k_{i,j} \left(1 - \frac{l_{i,j}}{p_{i,j}}\right) \times y_i\right)^2};$$

**End Foreach**

/∗ Select top K nodes which have high value of energy remaining in DM.∗/
**Sort** DM in descending order;
$DM \leftarrow DM[0..\text{top} - K];$

/∗ Update energy function by Newton − Raphson method.∗/
**For** $i \leftarrow 1$ **To** $\frac{\text{count}(V)}{K}$ **Do**
 **Foreach** $v \in DM$ **Do**
  **Compute** $\delta x$ and $\delta y$ for v;
  $x_v \leftarrow x_v + \delta x;$/∗ Update the x coordinate of v.∗/
  $y_v \leftarrow y_v + \delta y;$/∗ Update the y coordinate of v.∗/

  /∗ Recalculate energy function E after position update.∗/
  $$E \leftarrow \sum_i \sum_j \frac{1}{2} w_i \times k_{i,j}\left(p_{i,j}{}^2 + d_{i,j}{}^2 - 2 \times d_{i,j} \times p_{i,j}\right)$$
 **End Foreach**
**End For**

Algorithm 2 Pseudo code of Batch Weight Updating algorithm

### 4.2 Weight determination and assignment

  Recall that in BWU algorithm, an attribute called "weight" is assigned to every node in the network topology to influence the placement strategy. We also stated that the KK algorithm updates the visual positions of nodes iteratively. The smaller the weight, the smaller the chance the node will be selected for visual position updating. On the contrary, the larger the weight, the higher the chance the node will be selected for visual position updating. Therefore, weights play a major role in node selection.

  The pseudo code of CalWeight algorithm is described in Algorithm 3. In CalWeight algorithm, we define five scaling factors $\delta_1, \delta_2, \delta_3, \delta_4, \delta_5$ for managing the weights. Scaling factors are real numbers between 0 and 1 and they are updated during the execution. The objective of these scaling factors is to determine the weight of the nodes. $\delta_1$ denotes the maximum weight of the nodes which visual position are supposed to be updated immediately on the following iteration. For the purpose of this discussion, we label these nodes as "active" nodes. $\delta_2$ specifies the maximum weight of active 1-hop nodes. $\delta_3$ specifies the maximum weight of active 2-hop nodes. $\delta_4$ specifies maximum weight of nodes which visual positions are either likely to be updated slightly or to remain unchanged. Finally, $\delta_5$ specifies the maximum weight of nodes which are considered to be anchor points.

  For instance, nodes assigned with $\delta_5$ may not require updating of visual positions during the execution of the algorithm. In that case, the weight of these nodes will carry a smaller value. On the contrary, the weight of "active" nodes will carry a higher value. Suppose that $\delta_1 = 0.98$ and a node $x$ is an "active "node. Also assume that the weight of node $x$ is 0.9 at 100th iteration. In this case, the new weight of node $x$ is $= \textit{value of weight} \times \delta_1 = 0.9 \times 0.98 = 0.882$. If the weight of node $x$ is still considered to be high after multiplying with a scaling factor, it will be selected again for visual position update in 101th iteration.

In addition, the CalWeight algorithm requires four inputs: (a) the network topology, (b) an array of anchor points (if any), (c) percentage of node for selection, and (d) the iteration of the first stage of main algorithm. CalWeight algorithm outputs a matrix to store the weights of nodes for BWU algorithm. CalWeight algorithm iteratively updates the weights of nodes by using a loop. On initialization, a node *sn* (starting node) with high weight value is selected by CalWeight algorithm. CalWeight algorithm then updates the weights of nearby nodes (i.e. 1-hop nodes of *sn*, 2-hop nodes of *sn* and so on) iteratively.

The loop of CalWeight is terminated when the *expcount* is larger than the iteration count (denoted by the *counter* variable) in the first stage of main algorithm. The lower bound of *expcount* is 1 and the upper bound of *expcount* is $\frac{1}{2}n$ where $n$ is the node count of network topology. These bounds specify the minimum and maximum number of nodes for weight updating on each iteration. The variable *counter* is initialized with a small value and is less than the *expcount* at the early stage of the processing. Therefore, a large number of nodes will be subjected to weight updating at the beginning. When the value of *counter* is incremented gradually, fewer nodes will selected for weight updating except for those nodes which still carry high weight values.

**ALGORITHM 3.** CalWeight algorithm

**Input**: a netwotk topology $\mathbf{G}(V, E)$
an array containing anchor points **AP**
percentage of node for selection **P**
iteration count of the main algorithm **counter**

**Output**: a matrix of weights **w**

**Initialize** the control varialble of expansion **expcount** ← 1;
**Initialize** the array of weight **w** ← ∅;

**Repeat** until expcount > counter
    **Initialize** the starting node **sn** ← random selection of nodes with high energy;
    **For** i ← 1 **To** expcount **Do**
        v1 ← 1 hop nodes of sn;
        v2 ← 2 hop nodes of sn;
        w[1..n] ← w[1..n] × $\delta_4$;
        w[v2] ← w[v2] × $\delta_3$;
        w[v1] ← w[v1] × $\delta_2$;
        w[sn] ← w[v1] × $\delta_1$;
        w[AP] ← w[AP] × $\delta_5$;
        **If** |v1| > |v2| **Then** {
            sn ← v1;
        **Else**
            sn ← v2;
        **End If**
    **End For**
    expcount ← expcount + $\max\left(\frac{|V|}{2}, P \times (|v1| + |v2|)\right)$;
**End Repeat**
**return** w;

Algorithm 3 Pseudo code of CalWeight algorithm

**4.3 Estimation of folded regions**

The EstRegion algorithm for folded region estimation aims to discover regions in the visual drawing where there is significant variation in edge distribution in comparison to other regions. The pseudocode of EetRegion is given in Algorithm 4. The algorithm accepts (a) the estimated length of all edges (represented by signal strength), (b) the calculated length of all edges in visual drawing returned by BWU algorithm, and (c) a threshold for folded regions estimation. The EstRegion algorithm produces a subset of network topologies (nodes and edges) which is significantly different from others.

First, the algorithm initializes the variable $diff$ for each edge in the network topology. The value $diff$ is used to determine the difference between the length of edges which is estimated from the signal strength and the length of edges in visual drawing returned by BWU algorithm. Second, EstRegion algorithm calculates the variable $stdev$ which is the standard deviation of the $diff$. Finally, EstRegion algorithm calculates the ratio $r$ for every edge, which specifies the variation in the length of edges in the visual drawing. The calculation of approximation error $r$ is shown in following equation:

$$r(x) = \frac{diff_x - avg(diff)}{stdev} \tag{6}$$

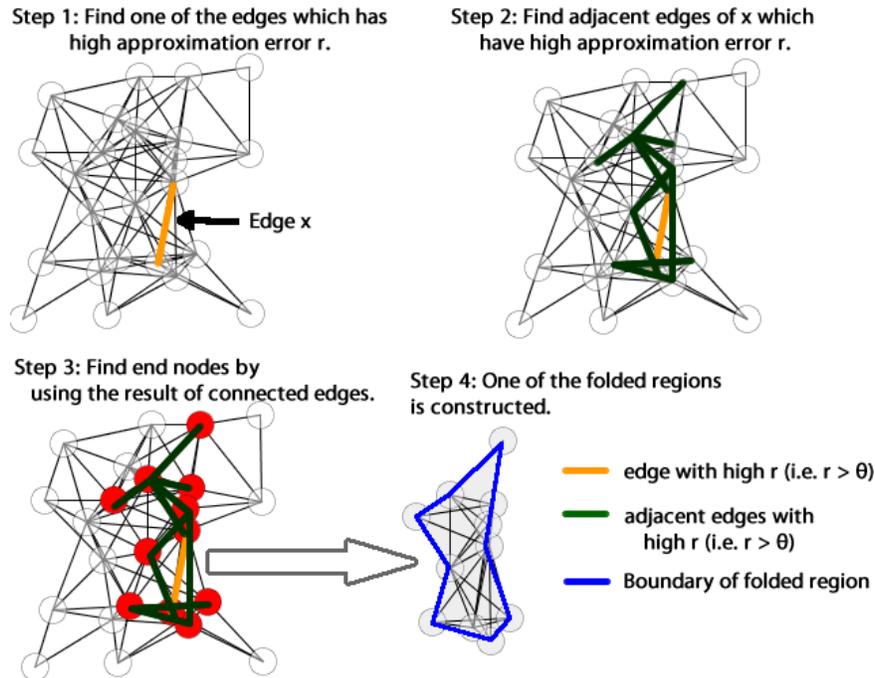

Figure 4 Step by step construction of a folded region by the EstRegion algorithm

A folded region is a set of edges situated near each other which having large approximation error $r$. The procedure of folded region construction will be executed after the approximation error of edges in the network topology has been calculated. The folded region construction procedure can be described in the following steps. Firstly, EstRegion algorithm compares the approximation error $r$ of edges in the network topology with the threshold $\theta$. For instance, if the approximation error $r$ of an edge $x$ is less than $\theta$, then the edge is omitted. Otherwise, EstRegion algorithm finds all its adjacent edges which have high approximation error $r$. This can be achieved by searching 1-hop edges of the chosen edge $x$, 2-hop edges of $x$, and so on. Searching continues until no high approximation error $r$ of edges can be found.

Secondly, after the searching is done, EstRegion algorithm counts the total number of edges with high approximation error $r$ found in the previous step. The algorithm will omit the result of searching if the number of edges with high approximation error $r$ is less than 10. The reason for omitting is that if the numbers of edges are too small, the algorithm cannot unfold these small regions effectively.

Finally, if sufficient number of edges were found in the previous step, EstRegion algorithm then constructs a folded region by connecting the end nodes of edges which have high approximation error $r$. The end nodes are the nodes belong to an edge. In other words, the construction of folded regions use the result of edge searching described in previous steps. For example, the edge $x$ is from the result of edge searching and it has 5 end nodes. EstRegion algorithm will then append the edge $x$ and its 5 end nodes into a region. The same procedure is applied to other edges with high approximation error $r$. The algorithm repeats above steps iteratively until all edges in the network have been traversed. The pseudo code of EstRegion algorithm is given in Algorithm 4 and the steps of EstRegion algorithm are illustrated in Figure 4.

**ALGORITHM 4.**  EstRegion algorithm
---
**Input**: an array containing the length of edges in the network topology $G(V, E)$ $\mathbf{e_1}$
an array containing the length of edges in the visual drawing of $G(V, E)$ $\mathbf{e_2}$
the threshold for folded region estimation $\mathbf{\theta}$
**Output**: sets of nodes and edges in G which are significantly different from others

/∗ Calculate the value of diff for every edge in G.∗/
**Initialize** the array containing the value of difference of edges as **diff**;
**Foreach** e ∈ E **Do**
$$\text{diff}[e] \leftarrow \frac{e_1[e]}{\text{average}(e_1)} - \frac{e_2[e]}{\text{average}(e_2)};$$
**End Foreach**

/∗ Calculate the standard deviation of edges.∗/
**Initialize** stdev $\leftarrow \sum_{e=1}^{n} \sqrt{\left(\text{diff}[e] - \text{average}(\text{diff})\right)^2}$ ;

/∗ Calculate the approximation error for every edge in G.∗/
**Foreach** e ∈ E **Do**
$$\text{initialize } r[e] \leftarrow \frac{\text{diff}[e] - \text{average}(\text{diff})}{\text{stdev}};$$
**End Foreach**

/∗ Construct folded regions by using the value of r.∗/
**Initialize** folded regions as $\mathbf{R} \leftarrow \emptyset$;
**Initialize** the count of folded regions as **count** ← 0;
**For** i ← 1 **To** count(r) **Do**
  /∗ Omit edges which the value of r is less than threshold θ.∗/
  **If** r[i] < $\theta$ **Then**
    **Continue**;
  **End If**

```
        /∗ Find a set of adjancent edges with high r.∗/
        Initialize edges with high r as E_ADJ ← r[i];
        For j ← 1 To count(adjancent edges of r[i]) Do
                If r[j] ≥ θ Then
                        E_ADJ ← r[j];
                End If
        End For

        /∗ Omit the result once the count of edges is small.∗/
        If count(E_ADJ) < 10 Then
                Continue;
        End If

        /∗ Construct a folded region by connecting end nodes of E_ADJ.∗/
        Foreach v ∈ V Do
                Foreach e ∈ E_ADJ Do
                        If v is the end point of e Then
                                R ← v;
                        End If
                End Foreach
        End Foreach
        count ← count + 1;
End For

return R;
```

Algorithm 4 Pseudo code of EstRegion algorithm

### 5. ELNET SYSTEM FOR ANALYZING CNCAH NETWORKS

This section introduces the design and implementation of the prototype system called ELnet for CNCAH networks. The following functions are implemented in ELnet system: (a) generation of syntactic CNCAH or common network topologies, (b) editing of existing network topologies, (c) execution of force-directed algorithms (e.g. Fruchterman Reingold, Davidson Harel, W-KK-MS, etc.), and (d) visualization and analysis of force-directed algorithms' outcomes. Figure 5 illustrates the main screen of ELnet system.

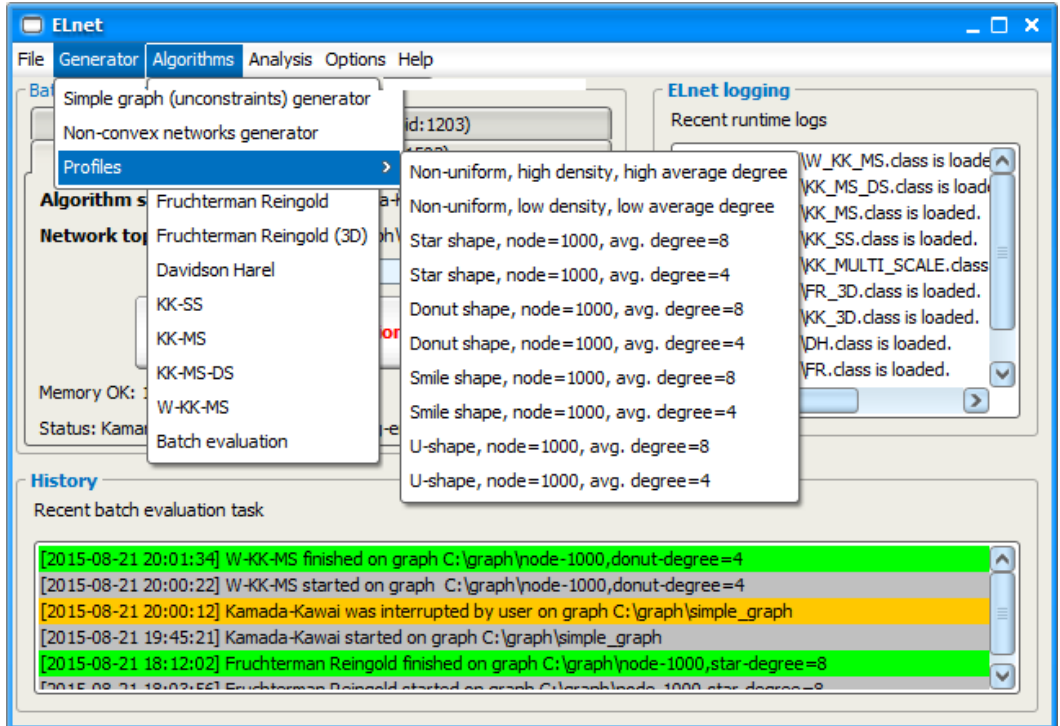

Figure 5 The main menu of ELnet

**5.1 Topology Generation**

Creating a sizable number of syntactic network topologies is essential for any experiments in mobile ad hoc network research. These generated network topologies are especially useful when we evaluate existing force directed algorithms for boundary nodes detection problem.

There are already promising applications that can be used for simple (random) graph generation. Examples of graph generation applications include NetworkX (Python programming language) [NetworkX developer team 2014], Open Graph Drawing Framework (C++ programming language) [Gutwenger et al. 2013], and JGraphT (Java programming language) [Naveh 2013]. Although the functions provided in these applications are sufficient for generating random and simple network topologies, they are not designed to generate CNCAH network topologies with relatively large number of control parameters. For example, CNCAH network topology generation process can be guided by the parameters such as number of nodes, average degree, distribution of edges and nodes, and distribution of clusters.

One of the key functions of ELnet is the bulk generation of simple networks as well as CNCAH networks by the use of a graphical interface. The network topology generator is able to generate instances of network topologies featuring different distributions of clusters, nodes, and edges. Figure 6(a) and (b) show samples of two network topologies with different distributions of clusters. Note that these two samples have the same number of nodes and the average degree.

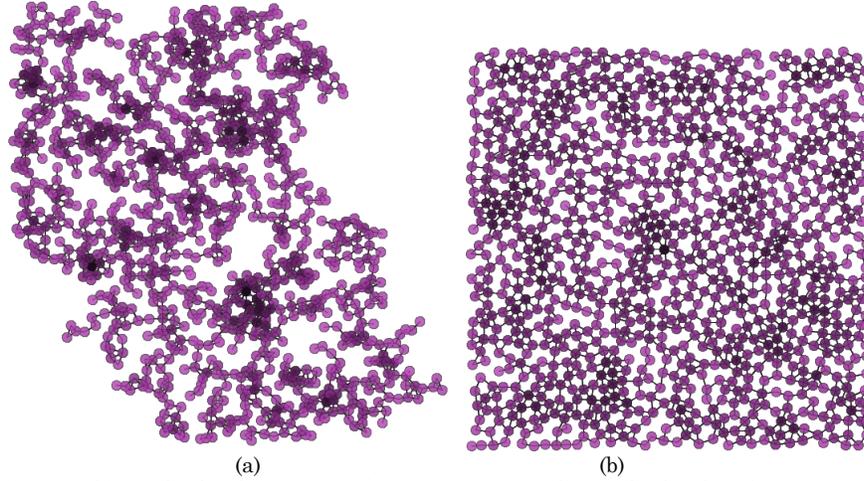
(a) (b)
Figure 6(a) Non-uniform distribution of network topology, (b) A uniform distribution of network topology.

A network topology consists of a finite number of nodes and edges. Recall that physical locations of the nodes are unknown for the CNCAH networks addressed in the paper. Therefore, a network topology could have infinite number of corresponding visual drawings. If we consider the case where a visual drawing is represented by pixels, the number of visual drawings could be infinite for different combination of screen sizes and resolutions. For example, the two possible sizes of a visual drawing for a given network topology could be either 640×480 or 1920×1080. Therefore, a unified coordinate system is necessary for generating network topologies especially for force-directed algorithms. In ELnet system, we adopt a 1 by 1 windowed coordinate system and all nodes are fitted into that coordinate system during the algorithm execution. After the execution of the algorithms is completed, users can visualize the results by selecting a desired size of visual drawing (e.g. 1920×1080) in ELnet. For the case of 1920x1080, the $x$ coordinates of nodes will be multiplied by 1920 and the $y$ coordinates of nodes will multiplied by 1080.

Screen captures of ELnet's topology generation methods are illustrated in Figure 10 and the parameters used by ELnet system in topology generation methods are summarized in Table 1. These parameters are used as the seeds for randomization.

Table 1 Parameters used in topology generation methods

| Parameter | Description |
|---|---|
| Number of nodes ($n$) | Expected number of nodes within the network topology |
| Average degree ($\delta$) | Expected average degree for the network topology |
| Node distribution ratio ($d$) | Minimum distance between nodes |
| Edge radius ($\gamma$) | Maximum distance between edges |
| Edge connectivity ($\gamma_b$) | The probability value used for generation of edges |
| Edge distribution ($e$) | The minimum distribution distance between edges |

The network topology generation algorithm consists of two stages; the node generation and edge generation. The first stage of the algorithm executes iteratively until the count of nodes exceeds the parameter ($n$). In every iteration, a newly generated node is assigned with a randomized $x$ and $y$ coordinates. The node will be accepted when it satisfies the parameters input by the user (e.g., $d$, $n$, etc). If the node does not satisfy one of the parameters, the node is discarded.

The second stage of the algorithm uses the parameter $\delta$ to determine how many edges should be generated randomly for the network topology. Edge randomization is performed iteratively by randomly selecting a node and generating the corresponding edges. Similar to the first stage, a generated edge will be accepted when it satisfies the parameters input by the user. Otherwise, the edge will be discarded. The loop for the generation of edges will terminate when the average degree of network topology exceeds the threshold ($\delta$) input by the user. The pseudo code of network topology generation algorithm is described in Algorithm 5.

---

**ALGORITHM 5.**   Network topology generation algorithm
---
**Input**: a windowed coordinate system **W**
   parameters of network topology generation **n, d, $\delta$, e, $\gamma$, $\gamma_b$**
**Output**: a network topology of **G** in windowed coordinate system
**Initialize** the output nodes $W_N \leftarrow \emptyset$;
**Initialize** the output edges $W_E \leftarrow \emptyset$;

/∗ First stage − Node generation.∗/
**While** count($W_N$) < n
   **Let** $v_R$ be a new node where $v_R \notin W_N$;
   $v_R.x \leftarrow$ A random x coordinate;
   $v_R.y \leftarrow$ A random y coordinate;
   **Foreach** $v \in W_N$ **Do**
      **If** distance($v_R, v$) < d **Then**
         $v_R.\text{reject} \leftarrow$ true;
      **End If**
      **If** $v_R.\text{reject}$ = false **Then**
         $W_N \leftarrow v_R$;
      **End If**
      $v_R.\text{reject} \leftarrow$ false;
   **End Foreach**
**End While**

/∗ Second stage − Edge generation.∗/
**While** degree($W_E$) < $\delta$
   **Foreach** $v \in W_N$ **Do**
      **Let** $v_R \in W_N$ where $v_R \neq v$;
      **Initialize p** $\leftarrow$ a random number between 0 and 1;
      **If** p $\leq 1 - \gamma_b$ **Then**
         $v_R.\text{reject} \leftarrow$ true;
      **Else If** distance($v_R, v$) < e **Then**
         $v_R.\text{reject} \leftarrow$ true;
      **Else If** distance($v_R, v$) > $\gamma$ **Then**
         $v_R.\text{reject} \leftarrow$ true;
      **End If**
      **If** $v_R.\text{reject}$ = false **Then**
         $W_E \leftarrow \langle v_R, v \rangle$;
      **End If**
      $v_R.\text{reject} \leftarrow$ false;
   **End Foreach**
**End While**

Algorithm 5 Pseudo code of network topology generation

ELnet also provides several visualization features on manipulating network topologies. For example, users can zoom in and zoom out on various network topologies. Users can also change the color of nodes which have high average degree. Users can also hide or display the nodes or the edges during the visualization. These features are shown in Figure 7(a) and Figure 7(b).

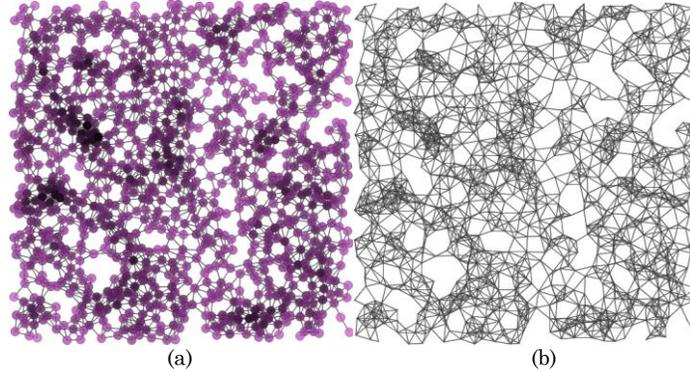

(a)          (b)

Figure 7(a) An enhanced view of the nodes in a network topology, (b) An enhanced view of the edges in a network topology.

### 5.2 Editing of existing network topologies

In ELnet system, users can also construct complex network topologies based on existing network topologies. For instance, users are able to alter existing network topologies and construct variations of these network topologies. Users can also add or remove nodes and edges from existing network topologies. New nodes and edges can also be generated within a selected region of the network. Figure 8(a) and Figure 8(b) show examples of generating a network topology and modifying it later.

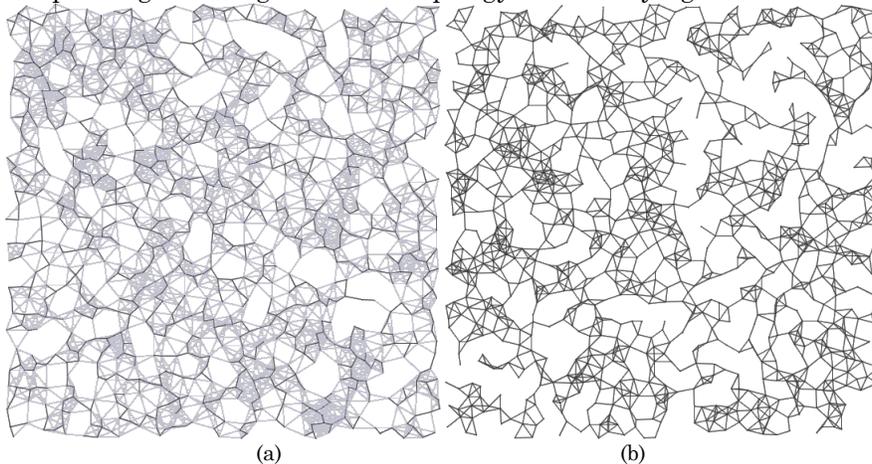

(a)          (b)

Figure 8(a) A visualization of a network topology, (b) After removing certain nodes and edges from (a).

Through the interface, users can also alter parameters as shown in Table 1 and produce a new network instantly without regenerating the entire network topology. Specifically, users can fine-tune an existing network topology or dynamically build the different network topologies from a single source. Moreover, ELnet can perform clipping operations for existing network topologies. The graphical user interface of ELnet provides features which can be used to generate complex and irregular shapes such as u-shape, circle, star, etc. Figure 9 illustrates two irregular shapes network topologies generated by ELnet. Figure 10 illustrates a screenshot of ELnet for generating an irregular shaped network topology through a user interface.

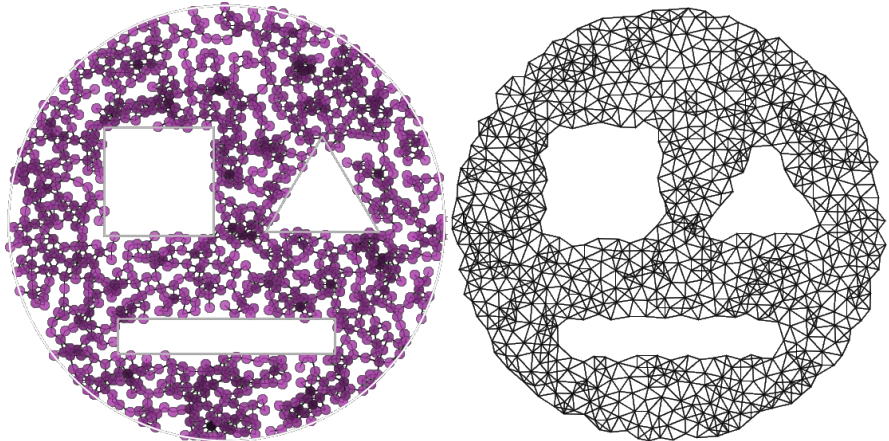

Figure 9 Examples of irregular network topologies.

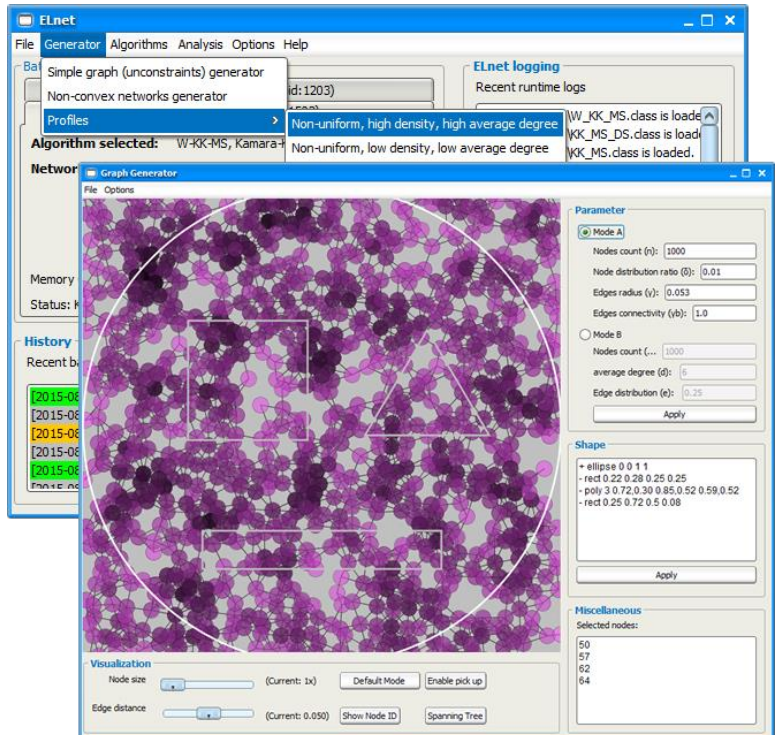

Figure 10 The screen capture of network topology generator.

During the network topology generation, the number of nodes, the distribution of nodes and edges, and the average degree of network topology can affect the structure of the network topology. Node distribution ratio ($d$) and edge distribution ratio ($e$) from Table 1 can be used to control the proximity of the nodes and edges. The lower these values, the higher the chance that the nodes will stick to each other. The higher the value, the more sparse the edges and nodes will be. By adjusting these distribution ratios, it is easy to modify an existing network into a sparse or an uniformly distributed topology. In addition, by assigning a lower value to the node distribution ratio ($d$) and maintaining the value of the edge distribution ratio ($e$) and edge radius ($\gamma$), users can construct clusters within each network topology. The reason for this is that the lower the node distribution ratio ($d$), the higher the density of the nodes in some regions of the network topology. By balancing these parameters, we can obtain sparse or uniform distribution of clusters.

For testing CNCAH networks, ELnet provides a feature that allows users to generate network topologies with irregular shapes. The system can construct network topologies with polygons. The following is a sample script used to generate a network topology with an irregular shape. The result of the script is shown in Figure 9.

$+ ellipse$ 0 0 1 1
$- rect$ 0.22 0.28 0.25 0.25
$- poly$ 3 0.72,0.30 0.85,0.52 0.59,0.52
$- rect$ 0.25 0.72 0.5 0.08

The coordinates used in the script are within a 1 by 1 windowed coordinate system. The + and – signs in the script specify the areas where nodes and edges can be generated. Nodes and edges can only be generated in open areas which are denoted by + label. Nodes and edges cannot be generated in the closed areas denoted by – labels. Moreover, ELnet supports three kinds of polygons: ellipse, rectangle (rect) and polygon (poly).

- Ellipse requires four parameters: the first two are the x, y coordinates of left-top corner and the last two are the width and height of the ellipse.
- A rectangle can be defined with four parameters: the x, y coordinates of left-top corner and the x, y coordinates of the right-bottom corner.
- A polygon can be specified with at least four parameters: the number of points (first parameter) and the coordinates of at least three points.

**5.3 True boundary node identification in a visual graph**

A boundary node identification algorithm is implemented in ELnet system. The algorithm identifies boundary nodes in a visual graph where screen positions (x and y coordinates) of the nodes are known. The purpose of this algorithm is to identify the true boundary nodes so that they can be compared with the boundary nodes detected by the forced directed algorithms such as KK which solely relies on the network topology information.

In plane geometry, a plane consists of nodes and edges. An atomic plane is a plane which does not embed any other planes or sub-planes. In ELnet system, the visual graph for the network topology is first projected onto a two-dimensional drawing frame before executing the boundary node identification algorithm. Outward boundary is the plane in the two-dimensional drawing frame which does not intersect and embed any other sub-planes. Therefore, the largest intersection-free plane of the two-dimensional drawing frame is the outward boundary. That is, nodes on the largest plane are considered to be the boundary nodes. Figure 11(a) illustrates all possible planes of a two-dimensional drawing frame consisting nodes A, B, C, D, E, and F. The three planes in the polygon are $ABCF$ (atomic plane), $ABCDEF$ and $CDEF$ (atomic plane). The largest plane is $ABCDEF$ and it is also the boundary of the two-dimensional drawing frame. The time complexity for the node identification algorithm is $O(|N| * |E|)$ where $N$ and $E$ are the node count and the edge count. The pseudo code of boundary node identification is given in Algorithm 6.

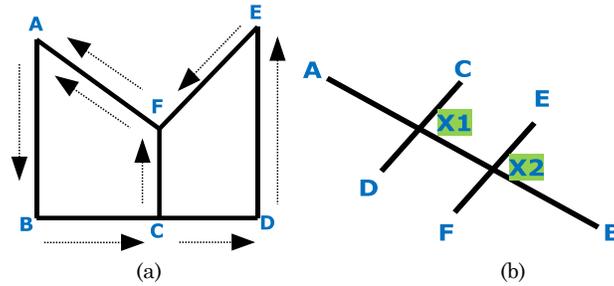
(a) (b)
Figure 11(a) An example of planes, (b) Adding two dummy node to crossing edges.

The algorithm performs following steps for boundary node identification.
1. Determine the clockwise or counterclockwise ordering of the edges.
2. Separate all crossing edges by adding dummy nodes (X1 and X2) as shown in Figure 11 (b).
3. Calculate the angle of the edges on each node in order to traverse the planes (either clockwise or counterclockwise).
4. Select a starting edge by choosing a node located at the corner and picking one of the edges connected to the node.
5. Start traversing the two-dimensional plane from the selected edge in counterclockwise or clockwise direction until returning to the selected edge (i.e. all edges and nodes have been traversed and all the planes have been found).
6. Merge the planes found in the previous step and eliminate any dummy nodes.
7. Nodes on the largest plane are the boundary nodes.

**ALGORITHM 6.**   True boundary nodes identification algorithm
**Input**: a visual drawing of network topology **G**
**Output**: boundary nodes of **G**

/∗ Check whether PARENT and CHILDREN are polygons.∗/
**Def Function** is_sub_planes(PARENT, CHILDREN)
    **Foreach** child ∈ CHILDREN **Do**
        **If** child ∈ PARENT **Then**
            **return** true;
        **End If**
        **Foreach** v ∈ child **Do** /∗ v is a node of child.∗/
            **If** v **is a node of** PARENT **Then**
                **return** true;
            **End If**
        **End Foreach**
    **End Foreach**
    **return** false;
**End Def Function**

/∗ Find a plane starting from v1 and end at v2.∗/
**Def Function** travel_plane(v1, v2)
        **Initialize** the result of planes found as **RET** ← ∅;
        **Initialize** the starting node as $v_{START}$ ← v1;
        **While** $v_{START}$ <> v2
            result ← v1;

```
                result ← v2;
                Initialize the array containing the neighbors of v2 as v_NEIGHBOR;
                index ← indexOf(v_NEIGHBOR, v1);
                v1 ← v2;
                v2 ← v_NEIGHBOR[index + 1];
                result ← v2;
        End While
        ret ← Node_start;
        return ret;
End Def Function

/* Find the neighbors of nodes of v that v are nodes in G.*/
Initialize the array containing neighbors of node v where v ∈ G as neigh;
Foreach v ∈ G Do
        neigh[v] ← neighbors of v in counterclockwise ordering;
End Foreach

/* Traverse G and find planes in G.*/
Initialize the result of planes found in G as planes;
Initialize the node at the corner of G as v_CORNER;
Foreach v ∈ G Do
        For i ← 1 To count(neigh) Do
                planes ← travel_plane(v_CORNER, neigh[v][i]);
        End For
End Foreach

/* Remove duplicated planes.*/
Foreach p1 ∈ planes Do {
        Foreach p2 ∈ planes Do
                If p1 <> p2 and is_sub_planess(p1, p2) = true Then
                        planes ← delete p2;
                End If
        End Foreach
End Foreach

/* Identify boundary nodes in planes.*/
Initialize the array containing the boundary nodes of G as B;
Foreach p ∈ planes Do
        Foreach v ∈ p Do /* v is a node in the plane p.*/
                B ← v;
        End Foreach
End Foreach

return B;
```
Algorithm 6 Pseudo code of true boundary nodes identification algorithm

### 5.4 Execution of force-directed algorithms

Classical force-directed algorithms (e.g. Kamada-Kawai, Fruchterman Reingold, Davidson Harel, etc) and the proposed W-KK-MS algorithms were implemented in the proposed ELnet system. The screenshots of the event logs and algorithm selection are shown in Figure 12. In ELnet system, users can schedule the algorithms to be executed in batch. For example, users can select algorithms Kamada-Kawai, Fruchterman Reingold, Davidson Harel, and W-KK-MS to be tested on a u-shape network topology for evaluation or select other algorithms and network topologies where are saved in a local directory. Furthermore, users can set various termination requirements for the algorithms. The termination requirements implemented for all force-directed algorithms are as follows:

- Number of iterations
- Percentage of sensitivity
- Percentage of specificity
- Percentage of accuracy
- Time limit

The ELnet system was developed in Java programming language and *generic programming* technology was used during the implementation process.

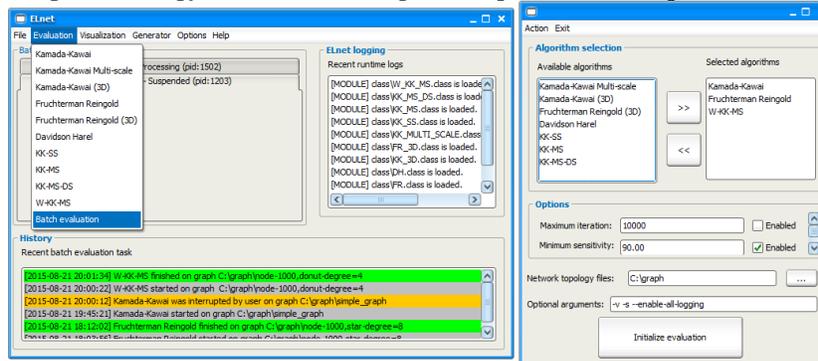

Figure 12 The screenshots of algorithm evaluation.

**5.5 Visualization and analysis of force-directed algorithms**

A logging mechanism is also implemented in the ELnet system. In ELnet system, the algorithm executions are logged and traceable. The traceable information which is recorded for every iteration of the algorithm includes:

1. The accuracy, sensitivity, and specificity.
2. The time spent.
3. The true positive, false negative, false positive, and true negative information of boundary node detection.
4. The visual drawing of network topology.
5. The nodes and edges information.

Elnet provides an analysis function based on traceable information. An example of visualization of experimental results and analysis of algorithms is shown in Figure 13.

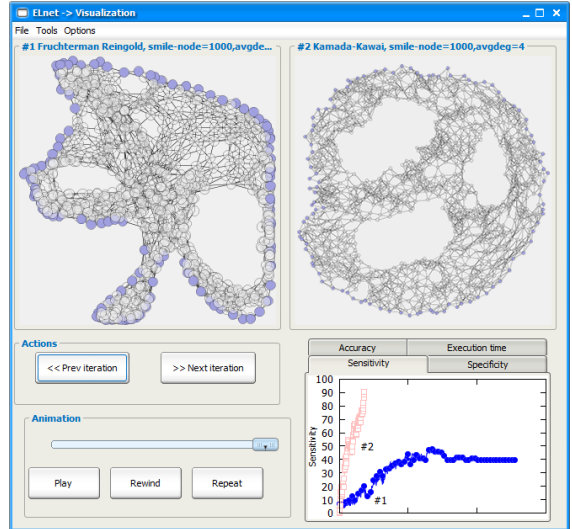

Figure 13 The screenshot of visualization of intermediate iterations and performance analyzing.

## 6. EXPERIMENTS

In this section, we evaluate the proposed W-KK-MS algorithm for CNCAH networks with Kamada-Kawai (KK), Fruchterman Reingold (FR) and Davidson Harel (DH). We also examine the visual drawings of the algorithms' output during its iteration process. These experiments were performed with a personal computer containing an Intel Pentium T2390 processor, 4GB of memory and Windows XP 32-bit.

By using ELnet system, we generated several benchmarks of CNCAH networks that had been adopted in previous studies [Efrat et al. 2010], [Saukh et al. 2010] for our experiments. Node counts and average degree of these networks are 1000 and 8 respectively. In these benchmark topologies, there are 4 kinds of irregular shapes. We used the ELnet system to generate five CNCAH networks per each irregular shape. In total, the experiment consisted of 20 instances of network topologies.

We set $\delta_1 = 1.0$, $\delta_2 = 0.95$, $\delta_3 = 0.7$, $\delta_4 = 0.05$ and $\delta_5 = 0$, $\theta = 4$ and $P = 0.1$ for the experiments. Figure 14 illustrates CNCAH networks generated for the experiment.

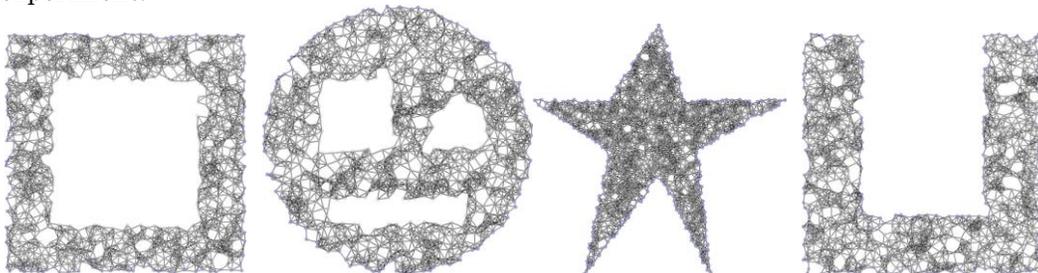

Figure 14 Samples of CNCAH networks generated for the experiments by the ELnet system.

In the experiments, we measure four types of performance metrics: the sensitivity, specificity, accuracy and execution time with respect to the varying number of nodes and different average degrees. The definition of them are described in Table 2.

Table 2 Performance evaluation metrics

| Metrics | Description | |
|---|---|---|
| Sensitivity/True positive rate | The percentage of boundary nodes on the initial network topology correctly identified as boundary nodes by algorithms | The higher the better |
| Specificity/True negative rate | The percentage of non-boundary nodes on the initial network topology correctly identified as non-boundary nodes by algorithms | The higher the better |
| Accuracy | The sum of the true positive count (i.e., boundary nodes correctly identified as boundary nodes) and the true negative count (i.e., non-boundary correctly identified as non-boundary nodes) divided by the total number of nodes examined | The higher the better |
| Execution time | Total amount of execution time that the algorithm ran in seconds | The lower the better |

Figure 15 illustrates the experiment settings to obtain the sensitivity, specificity and accuracy of W-KK-MS algorithm by using the Algorithm 6 given in section 5.3.

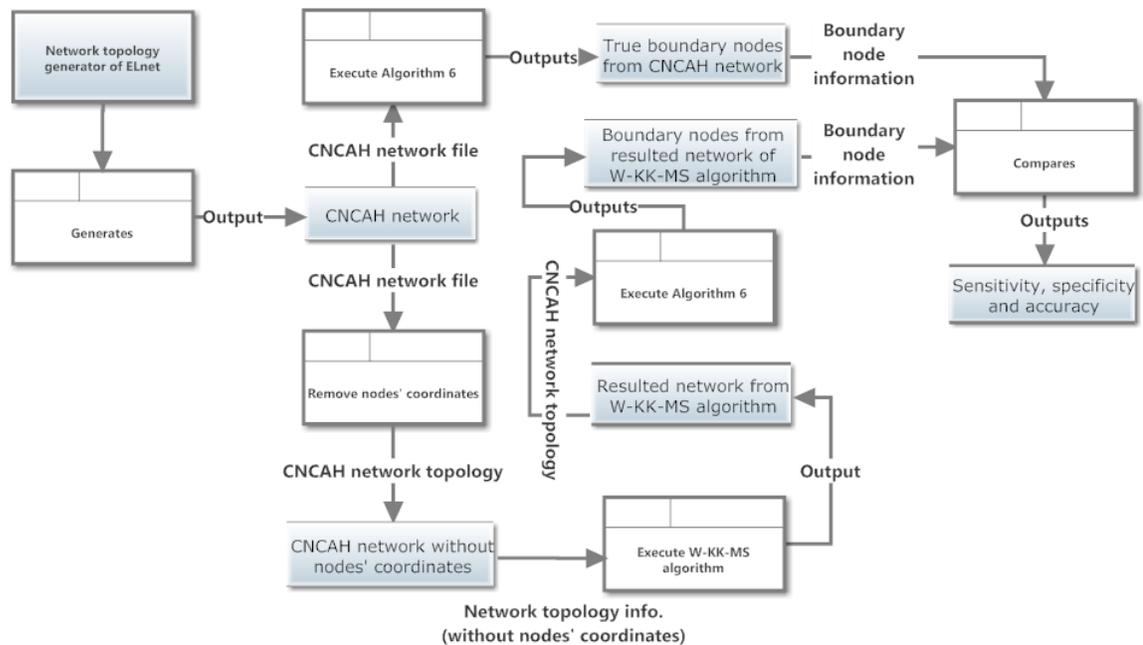

Figure 15 Experiment settings

## 6.1 Evaluation of execution time for detecting boundary nodes and unfolding

In this section, we compare the total time spent by all algorithms for achieving 90% sensitivity for CNCAH networks. Two stopping criteria were set for this experiment. The algorithm will stop when it either archives 90% sensitivity or the sensitivity of algorithm remains unchanged up to 100 iterations.

Figure 16 summarizes the results of the evaluation. The horizontal axis in Figure 16 is used to denote the average execution time for each shape. From the experiments, we found that FR and DH could not reach 90% sensitivity. Although FR had a faster converge rate compared to KK and DH, but FR could not improve the sensitivity any further. The average sensitivity of FR is approximately in the range of 50% to 60% for all network topologies we have evaluated. We also found that DH could not reach 90% sensitivity and it does not significantly improve the sensitivity. It is the reason why the execution time of DH is flat on all experiments. KK is able to reach 90% sensitivity in experiments and it spent approximately 1000 to 3800 time units to reach that level. From the experimental results, we can see that the proposed W-KK-MS is able to reach 90% sensitivity in the shortest time.

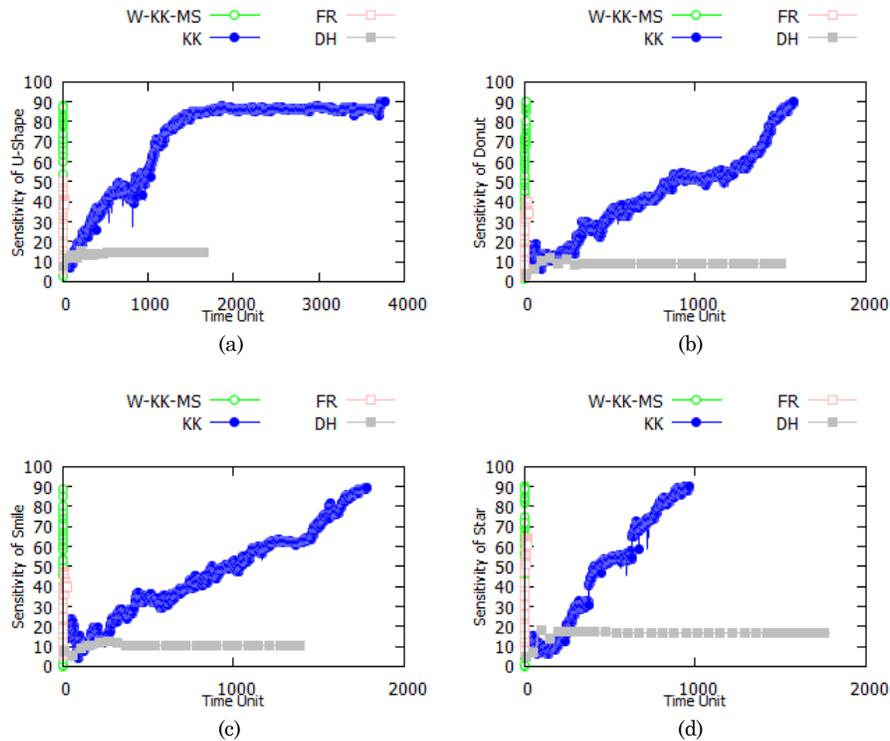

Figure 16 Evaluation of execution time.

A number of interesting results were also observed during the experiments. When we compared the final visual drawings of these irregular shapes, we found that the final visual drawing of the donut shape is similar to that of original shape except that it was rotated. Besides, we also found that in the final visual drawing of the smile shape, the algorithm could not properly display the polygons from the inner holes which consist of a triangle, a rectangle, and a square. The star shape and u-shape also had the same problem as the donut and smile shapes. They were rotated and the visual drawing was collapsed in the region where the nodes had low average degree. In the following sections, we compare the visualization of each shape at various time units when different algorithms are used.

### 6.1.1 Visualization of force-directed algorithms for star shape

Figure 17 depicts an intermediate visualization results of a star shape for four different algorithms. From the results shown in Figure 16(d), we can see that W-KK-MS took just 5 time units whereas KK took approximately 1000 time units to achieve 90% sensitivity for a star shape. The visualizations of the star shape at 90% sensitivity are depicted in the last column of Figure 17. We can also observe that FR and DH could not achieve 90% sensitivity within 1000 time units when they are compared to KK and W-KK-MS.

| Algorithm | Visualization at 1 time unit | Visualization at 3 time unit | Visualization at 100 time unit | Visualization at 90% sensitivity (Final result) |
|---|---|---|---|---|
| W-KK-MS | 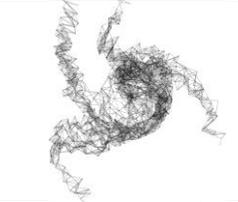 | 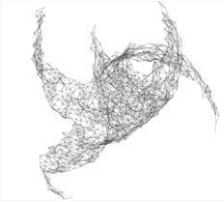 | 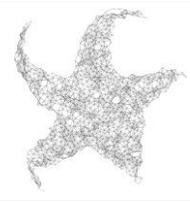 | 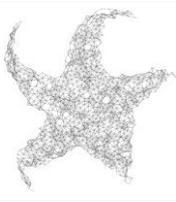 |
| KK | 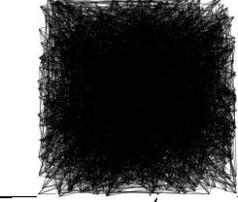 | 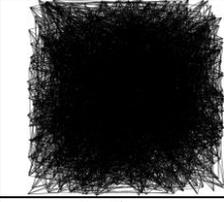 | 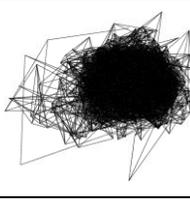 | 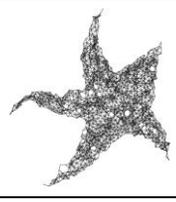 |
| FR | 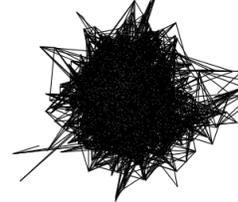 | 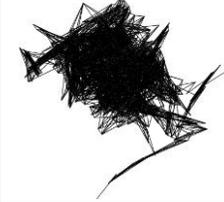 | 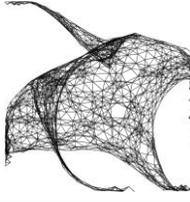 | Unable to reach 90% sensitivity |
| DH | 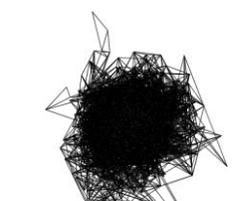 | 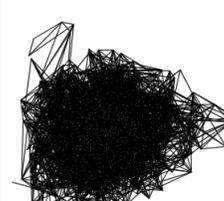 | 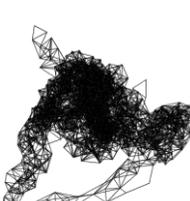 | Unable to reach 90% sensitivity |

Figure 17 (Column #2) star shape on 1 time unit, (Column #3) star shape on 3 time units, (Column #4) star shape on 100 time units, and (Column #5) star shape on 90% sensitivity and specificity.

### 6.1.2 Visualization of force-directed algorithms for u-shape

Figure 18 depicts an intermediate results of a u-shape visualization. According to the evaluation result reported in Figure 16(a), W-KK-MS took just 6 time units whereas KK took approximately 3800 time units to achieve 90% sensitivity for a u-shape. Similar to the previous case, FR and DH also cannot reach 90% sensitivity for the u-shape.

| Algorithm | Visualization at 1 time unit | Visualization at 2 time unit | Visualization at 100 time unit | Visualization at 90% sensitivity (Final result) |
|---|---|---|---|---|

| Algorithm | | | | |
|---|---|---|---|---|
| W-KK-MS | 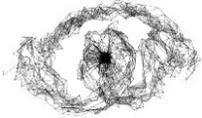 | 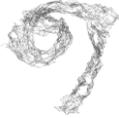 | 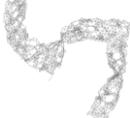 | 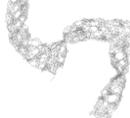 |
| KK | 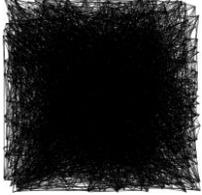 | 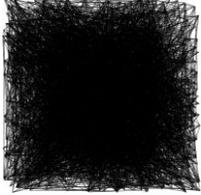 | 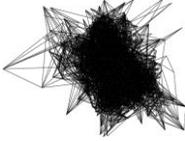 | 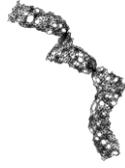 |
| FR | 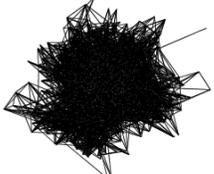 | 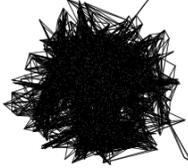 | 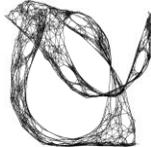 | Unable to reach 90% sensitivity |
| DH | 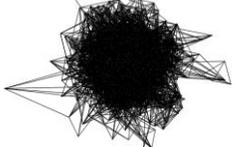 | 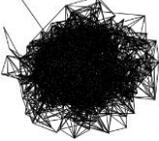 | 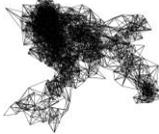 | Unable to reach 90% sensitivity |

Figure 18 (Column #2) u-shape results for 1 time unit, (Column #3) u-shape results for 2 time units, (Column #4) u-shape results for 100 time units and (Column #5) u-shape results for 90% sensitivity and specificity.

### 6.1.3 Visualization of force-directed algorithms for smile shape

Figure 19 depicts the intermediate results of a smile shape. According to the evaluation result reported in Figure 16(c), W-KK-MS took just 4 time units and KK took approximately 1800 time units to achieve 90% sensitivity for a smile shape. Similar to the previous cases, FR and DH also cannot reach 90% sensitivity for the smile shape when they are compared to KK and W-KK-MS.

| Algorithm | Visualization at 1 time unit | Visualization at 2 time unit | Visualization at 100 time unit | Visualization at 90% sensitivity (Final result) |
|---|---|---|---|---|
| W-KK-MS | 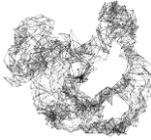 | 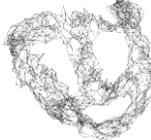 | 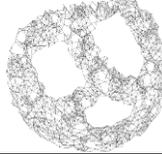 | 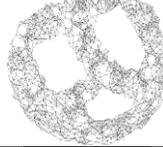 |
| KK | 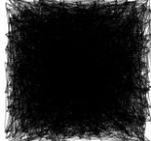 | 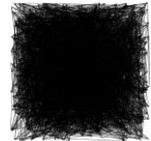 | 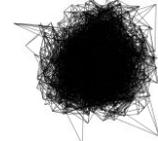 | 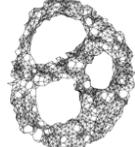 |
| FR | 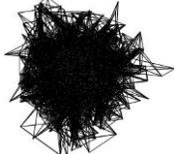 | 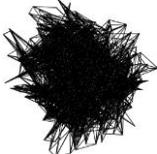 | 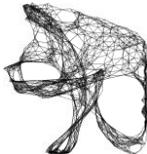 | Unable to reach 90% sensitivity |
| DH | 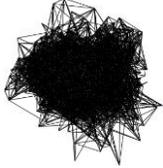 | 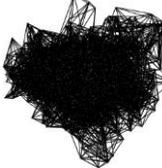 | 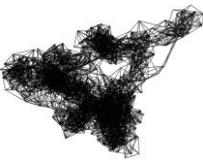 | Unable to reach 90% sensitivity |

Figure 19 (Column #2) smile shape results for 1 time unit, (Column #3) smile shape results for 2 time units, (Column #4) smile shape results for 100 time units, (Column #5) smile shape results for 90% sensitivity and specificity.

### 6.1.4 Visualization of force-directed algorithms for donut shape

The intermediate results of a donut shape visualization is depicted in Figure 20. According to the evaluation result in Figure 16(b), W-KK-MS took just 13 time units and KK took approximately 1700 time units to achieve 90% sensitivity for the donut shape. Similar to the previous cases, FR and DH also cannot reach 90% sensitivity for donut shape when they are compared to KK and W-KK-MS.

| Algorithm | Visualization at 1 time unit | Visualization at 5 time unit | Visualization at 100 time unit | Visualization at 90% sensitivity (Final result) |
|---|---|---|---|---|
| W-KK-MS | 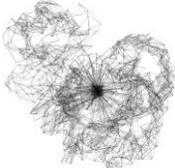 | 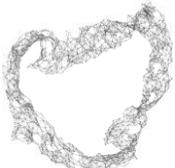 | 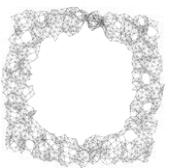 | 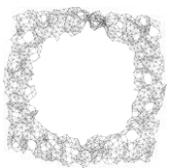 |
| KK | 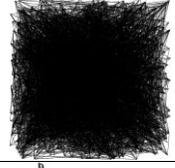 | 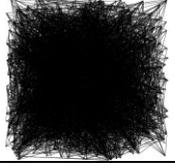 | 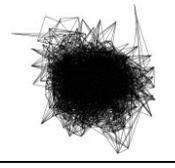 | 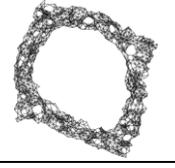 |
| FR | 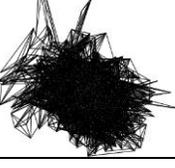 | 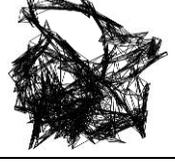 | 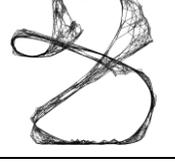 | Unable to reach 90% sensitivity |
| DH | 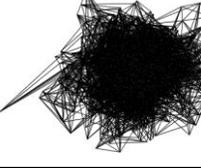 | 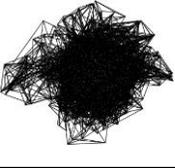 | 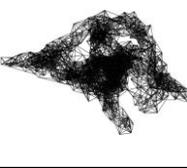 | Unable to reach 90% sensitivity |

Figure 20 (Column #2) Donut results for 1 time unit, (Column #3) Donut results for 5 time units, (Column #4) Donut results for 100 time units and (Column #5) Donut results for 90% sensitivity and specificity.

### 6.2 Evaluation of sensitivity, specificity and accuracy in boundary node detection

We also compared the sensitivity, specificity and accuracy of W-KK-MS for CNCAH networks with KK, FR, and DH. Two stopping criteria are used for this experiment. The algorithm will be stopped when it has either executed 60 time units or the value of the energy function equals to zero.

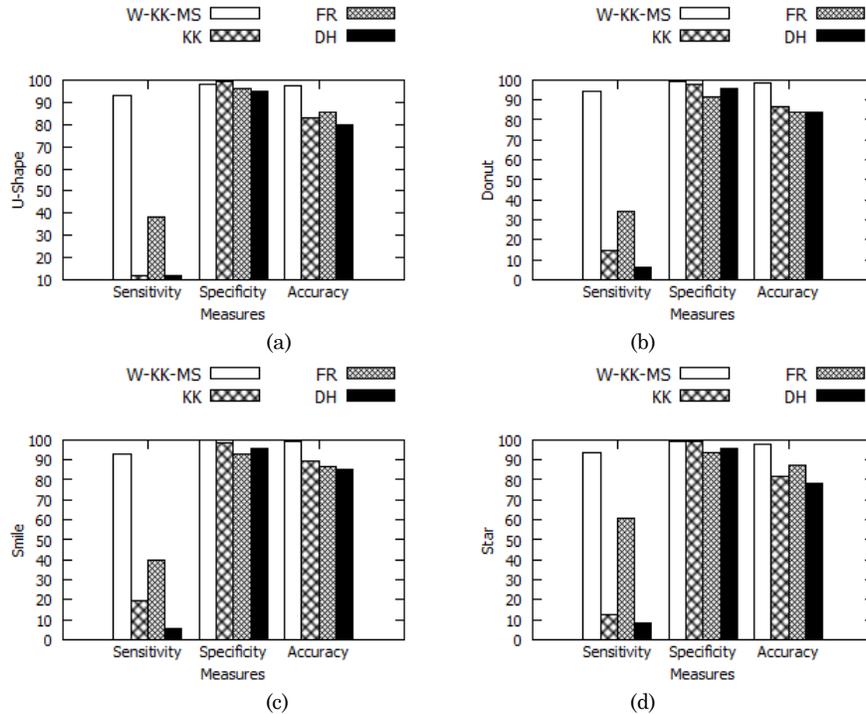

Figure 21(a) 60 time unit for U-Shape, (b) 60 time unit for Donut, (c) 60 time unit for Smile and (d) 60 time unit for Star.

Figure 21 shows the results of our evaluation verses different kinds of network topologies. According to the results of this evaluation, W-KK-MS algorithm achieved at least 93% in all three measures (sensitivity, specificity, and accuracy) for all the shapes tested in the experiment. It indicates that the performance of W-KK-MS algorithm was consistently high for W-KK-MS while some of the algorithms performed significantly lower. In particular, we can observe that W-KK-MS achieved the highest sensitivity and accuracy among all the algorithms compared. The result indicates that a large percentage of the boundary nodes have been identified except for a few nodes that are still distorted (especially around the corners).

Figure 22 shows the final visual drawings after each algorithm has been executed for 60 time units. These samples drawings were selected among other samples from the benchmark topologies.

| Algorithm \ Shape | W-KK-MS | Kamada-Kawai | Fruchterman Reingold | Davidson Harel |
|---|---|---|---|---|
| Donut | | | | |
| Smile | | | | |

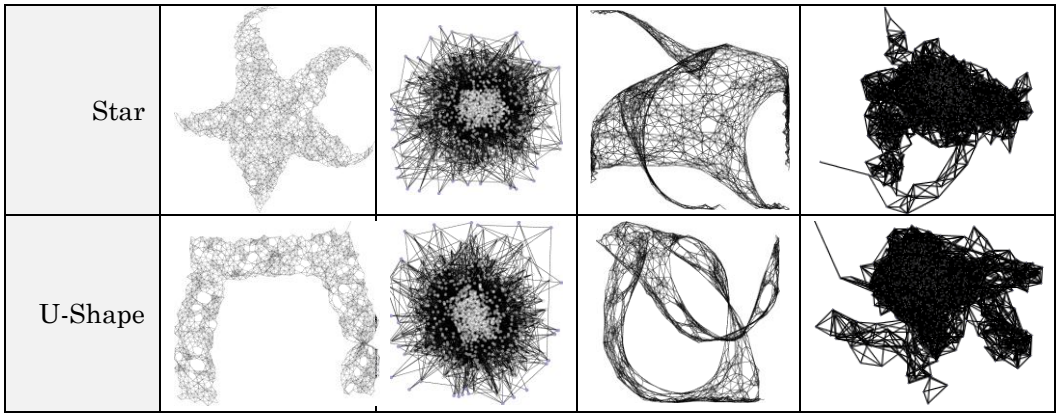

Figure 22 (Row #2) donut shape for 60 time units, (Row #3) smile shape for 60 time units, (Row #4) star shape for 60 time units and (Row #5) u-shape for 60 time units.

## 7. CONCLUSION

Our study targets the problem of using force-directed algorithms for detecting boundary nodes and visualizing CNCAH networks. The main contributions of our work is two-folds. First, we proposed a batch weight updating approach for KK in adjusting the node positions. The proposed algorithm called W-KK-MS is capable of discovering folded and twisted regions in CNCAH networks. To achieve fast convergence rate on boundary node detection, several approaches are proposed to control the node movements. An algorithm called EstRegion is also proposed to discover and repair possible folded and twisted regions. Experimental results show that W-KK-MS can achieve fast convergence on boundary detection in CNCAH networks and is able to successfully unfold stacked regions.

Second, the detailed design and implementation of a prototype system called ELnet for analyzing CNCAH networks is also presented in this paper. The primary purpose of system is to generate benchmark network topologies for testing, to enable the integration with third party algorithms, to visualize and analyze algorithms' performance. As part of the system, we also design an algorithm which can be used to identify boundary nodes from a visual drawing where the screen positions of the nodes are known. The output of this algorithm can be used to compare with the boundary nodes detected in topology-based algorithms such as KK, FR, DH, and the proposed W-KK-MS. The graphical interface of ELnet also provides features to visualize boundaries nodes on a visual drawing.

For the future work, we are planning to implement a distributed version of the proposed algorithm. Moreover, to improve the accuracy in constructing network topologies, we are planning to adopt technologies that are available for distance estimation over signal strength, e.g. RSSI (Received signal strength indication), and enhancement filtering, etc. [Volker et al. 2012]. We are planning to investigate the impact of noise and error measures in distance estimation since different hardware chipsets and environments could likely to produce variable results [Lui et al. 2011].

ACKNOWLEDGEMENT

This research was funded by the Research Committee of University of Macau, grant MYRG2017-00029-FST and MYRG2016-00148-FST.